\documentclass[a4,12pt]{article}
\usepackage[dvips]{graphicx,psfrag}
\def\dags{{\protect\mbox{\tiny \dag}}}
\def\sla#1{\mbox{$#1\hspace*{-0.17cm}\scriptstyle{/}\:$}}

\def\ls{\lower0.5ex\hbox{$\buildrel >\over{\scriptstyle\sim}$}} 
\def\rs{\lower0.5ex\hbox{$\buildrel <\over{\scriptstyle\sim}$}} 
\begin{document}
\pagestyle{empty} \setlength{\footskip}{2.0cm}
\setlength{\oddsidemargin}{0.5cm}
\setlength{\evensidemargin}{0.5cm}
\renewcommand{\thepage}{-- \arabic{page} --}
\def\mib#1{\mbox{\boldmath $#1$}}
\def\bra#1{\langle #1 |}  \def\ket#1{|#1\rangle}
\def\vev#1{\langle #1\rangle} \def\dps{\displaystyle}
%
% ------------------------------------------------------------
 \def\thebibliography#1{\centerline{REFERENCES}
 \list{[\arabic{enumi}]}{\settowidth\labelwidth{[#1]}\leftmargin
 \labelwidth\advance\leftmargin\labelsep\usecounter{enumi}}
 \def\newblock{\hskip .11em plus .33em minus -.07em}\sloppy
 \clubpenalty4000\widowpenalty4000\sfcode`\.=1000\relax}\let
 \endthebibliography=\endlist
 \def\sec#1{\addtocounter{section}{1}\section*{\hspace*{-0.72cm}
 \normalsize\bf\arabic{section}.$\;$#1}\vspace*{-0.3cm}}
 \def\subsec#1{\addtocounter{subsection}{1}\subsection*{\hspace*{-0.4cm}
 \normalsize\bf\arabic{section}.\arabic{subsection}.$\;$#1}\vspace*{-0.3cm}}
% ------------------------------------------------------------
% \vspace*{-1.7cm}
% \noindent
% \phantom
% {\large\bf July 16, 2007}

\vspace{-0.7cm}
\begin{flushright}
$\vcenter{
{  \hbox{{\footnotesize FUT-09-01}}  }
{  \hbox{{\footnotesize TOKUSHIMA Report}}  }
{  \hbox{(arXiv:0910.3049)}  }
%%% \phantom{  \hbox{{\footnotesize FUT-09-01}}  }
%%% \phantom{  {\hbox{{\footnotesize TOKUSHIMA Report}}}  }
%\phantom{\hbox{(arXiv:0910.3049)}}
}$
\end{flushright}

\vskip 0.8cm
\begin{center}
{\large\bf Search for anomalous top-gluon couplings at LHC revisited}
\end{center}

\vspace{0.6cm}
\begin{center}
\renewcommand{\thefootnote}{\alph{footnote})}
Zenr\=o HIOKI$^{\:1),\:}$\footnote{E-mail address:
\tt hioki@ias.tokushima-u.ac.jp}\ and\
Kazumasa OHKUMA$^{\:2),\:}$\footnote{E-mail address:
\tt ohkuma@fukui-ut.ac.jp}
\end{center}

\vspace*{0.4cm}
\centerline{\sl $1)$ Institute of Theoretical Physics,\
University of Tokushima}

\centerline{\sl Tokushima 770-8502, Japan}

\vskip 0.2cm
\centerline{\sl $2)$ Department of Information Science,\
Fukui University of Technology}
\centerline{\sl Fukui 910-8505, Japan}

\vspace*{2.8cm}
\centerline{ABSTRACT}

\vspace*{0.2cm}
\baselineskip=21pt plus 0.1pt minus 0.1pt
Through top-quark pair productions at LHC, we study possible effects of
nonstandard top-gluon couplings yielded by $SU(3)\times SU(2)\times U(1)$
invariant dimension-6 effective operators. We calculate the total cross
section and also some distributions for $pp\to t\bar{t}X$ as functions of
two anomalous-coupling parameters, i.e., the chromoelectric and chromomagnetic
moments of the top, which are constrained by the total cross section
$\sigma(p\bar{p} \to t\bar{t}X)$ measured at Tevatron. We find that LHC
might give us some chances to observe sizable effects induced by those new
couplings.

\vfill


\newpage
\renewcommand{\thefootnote}{$\sharp$\arabic{footnote}}
%-------------------------------------------------------------
\pagestyle{plain} \setcounter{footnote}{0}

% 111111111111111111111111111111111111111111111111111111111111
\sec{Introduction}

The Large Hadron Collider, LHC, now being about to operate \cite{LHC}, we will soon be
able to study physics beyond the standard model of the strong and electroweak interactions
in TeV world. Studies of such new physics can be classified into two categories: 
model-dependent and model-independent approaches. It is of course meaningless to
try to find which is more efficient: they have both advantage and disadvantage.
That is, the former could enable very precise calculations and analyses, but
we have to start again from the beginning if the wrong framework was chosen, while we
would rarely fail to get meaningful information in the latter but it would not be that
easy there to perform very precise analyses since we usually need to treat many
unknown parameters together. Therefore these two approaches to new physics should work
complementary to each other.

One reasonable way to decrease the number of such unknown parameters in a model-independent
analysis is to assume a new physics characterized by an energy scale ${\mit\Lambda}$
and write down $SU(3)\times SU(2)\times U(1)$-symmetric effective (non-renormalizable)
operators for the world below ${\mit\Lambda}$. Those operators with dimension 6 were
systematically listed in \cite{Buchmuller:1985jz}. Although we still have to treat
several operators (parameters) even in this framework, but some of the operators given
there were found to be dependent of each other through equations of motion
\cite{Grzadkowski:2003tf}. This shows that we might be further able to reduce the number of
independent operators, and indeed it was recently done in \cite{AguilarSaavedra:2008zc}.

In this effective-operator framework, not only electroweak couplings but also QCD couplings
receive nonstandard corrections. It will be hard for many readers to imagine that the QCD
couplings of light quarks are affected by those anomalous interactions, since the standard
QCD interaction form has so far been tested very well based on a lot of experimental data.
The top-quark couplings might however be exceptional, because this quark has not been studied
enough precisely yet, and its extremely heavy mass seems to tell us something about a new
physics beyond the standard model. That is, the $t$ quark could work as a valuable window
to a non-SM physics once LHC starts to give us fruitful data.

Under this consideration, we would like to perform here an analysis of anomalous top-gluon
couplings produced by the dimension-6 effective operators through top-quark pair productions
at LHC. We first describe our calculational framework in section 2. In section 3, we
calculate the total cross section of $p\bar{p}\to t\bar{t}X$ at Tevatron energy and compare
the result with the corresponding CDF/D0 data \cite{CDF-D0}, which gives a constraint on the
anomalous-coupling parameters. We then use them to compute the total cross section and also
some distributions for $pp \to t\bar{t}X$ at LHC, i.e., the top-angular, the
top-transverse-momentum, and the $t\bar{t}$-invariant-mass distributions. There we will find
that LHC might give us some chances to observe sizable effects induced by the new couplings.
Finally, a summary is given in section 4.

% 222222222222222222222222222222222222222222222222222222222222
\sec{Framework}

Let us clarify our basic framework in this section. In ref.\cite{Buchmuller:1985jz} were given
three effective operators contributing to strong interactions. Those operators produce top-pair 
production amplitudes which include $\gamma^\mu$, $\sigma^{\mu\nu}q_\nu$, $(p_i + p_j)^\mu$
and $q^\mu$ terms (or more complicated Lorentz structure), where $p_{i,j}$ and $q$ are
the top-quark $i$, $j$ and gluon momenta. However two of them were shown not to be
independent in \cite{AguilarSaavedra:2008zc}, and we only need to take into account
one operator
\begin{equation}
{\cal O}^{33}_{uG\phi}
=\sum_a [\:\bar{q}_{L3}(x)\lambda^a \sigma^{\mu\nu} u_{R3}(x)
\tilde{\phi}(x) G^a_{\mu\nu}(x)\:],
\end{equation}
where we followed the notation of \cite{AguilarSaavedra:2008zc}. This is quite a reduction.
Now the anomalous top-gluon couplings are given by
\begin{equation}
{\cal O}_{gt}
=\frac1{2\sqrt{2}} v \sum_a \bar{\psi}_t(x) \lambda^a \sigma^{\mu\nu} (1+\gamma_5) \psi_t(x)
G_{\mu\nu}^a(x),
\end{equation}
and our starting Lagrangian thereby becomes with unknown coefficients $C_{uG\phi}^{33}$ as
\begin{eqnarray}
&&\!\!\!\!{\cal L}
={\cal L}_{\rm SM} + \frac1{{\mit\Lambda}^2}
[\:C_{uG\phi}^{33}{\cal O}_{gt} + C_{uG\phi}^{33*}{\cal O}_{gt}^{\dags}\:] \nonumber\\
&&\!\!\!\!\phantom{{\cal L}}
={\cal L}_{\rm SM} + \frac1{\sqrt{2}{\mit\Lambda}^2}v
\sum_a[\:{\rm Re}(C_{uG\phi}^{33}) \bar{\psi}_t(x) \lambda^a \sigma^{\mu\nu} \psi_t(x) \nonumber
\\
&&\phantom{{\cal L}_{\rm SM} + \frac1{\sqrt{2}{\mit\Lambda}^2}v\sum_a[}
+ i\,{\rm Im}(C_{uG\phi}^{33}) \bar{\psi}_t(x) \lambda^a \sigma^{\mu\nu} \gamma_5 \psi_t(x)\:]
G_{\mu\nu}^a(x).
\label{Lag}
\end{eqnarray}
Here $v$ is the Higgs vacuum expectation value ($=246$ GeV), and ${\rm Re}(C_{uG\phi}^{33})$
and ${\rm Im}(C_{uG\phi}^{33})$ correspond to the top-quark chromomagnetic and chromoelectric
moments respectively.

As a matter of fact, a number of analyses including nonstandard couplings have been
performed in $t\bar{t}$ productions at high-energy hadron colliders ever since more than
a decade ago \cite{Haberl:1995ek,Atwood:1992vj}. However, the couplings used there were
not always the same. The precision of CDF/D0 data used there was not that
high either. In contrast to it, we can now state that the analysis using the two moments
is the most general model-independent one within the framework of effective operators.
Therefore, it must be worth to revisit the CDF/D0 data, to refine the
constraints on the anomalous couplings, and to apply the resultant information to
$pp \to t\bar{t}X$ at LHC, which is about to operate.

\begin{center}
\begin{figure}[htbp]
\begin{minipage}{14.8cm}
\hspace*{3.5cm}
{\includegraphics[width=7cm]{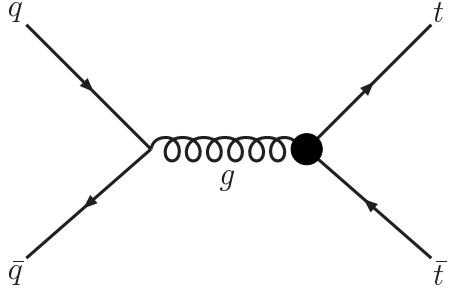}}
\caption{Feynman diagram of $q\bar{q} \to t\bar{t}$. The bullet $\bullet$ expresses
the vertex which includes the anomalous couplings.}\label{Feynqq}
\end{minipage}
\end{figure}
\end{center}

Apart from QCD higher order corrections, $q\bar{q}\to g\to t\bar{t}$ process is expressed
by one Feynman diagram (Fig.\ref{Feynqq}), and the corresponding invariant amplitude
is given by
\begin{eqnarray}
&&{\cal M}_{q\bar{q}}
=\frac1{4\hat{s}}g_s^2 \sum_a \bar{u}(\mib{p}_t) \lambda^a
{\mit\Gamma}^\mu(q) v(\mib{p}_{\bar{t}}) \,
\bar{v}(\mib{q}_2) \lambda^a \gamma_\mu u(\mib{q}_1),
\end{eqnarray}
where $q\equiv q_1 + q_2 (=p_t + p_{\bar{t}}),\: \hat{s}\equiv q^2$, $[a]$ is the color label
of the intermediate gluon,\footnote{Here (and hereafter) we do not show the color-component
    indices of $u$/$v$ spinors, and also all the spin variables for simplicity.}
we expressed the anomalous-coupling parameters as
\[
d_V = \frac{\sqrt{2}vm_t}{g_s{\mit\Lambda}^2} {\rm Re}(C^{33}_{uG\phi}),\ \ \ \
d_A = \frac{\sqrt{2}vm_t}{g_s{\mit\Lambda}^2} {\rm Im}(C^{33}_{uG\phi}),
\]
and we defined as
\[
{\mit\Gamma}^\mu(q) \equiv \gamma^\mu - \frac{2i\sigma^{\mu\nu}q_\nu}{m_t}
(d_V + id_A \gamma_5).
\]

\vskip 0.4cm

\begin{center}
\begin{figure}[htbp]
\begin{minipage}{14.8cm}
\hspace*{1.6cm}
{\includegraphics[width=11cm]{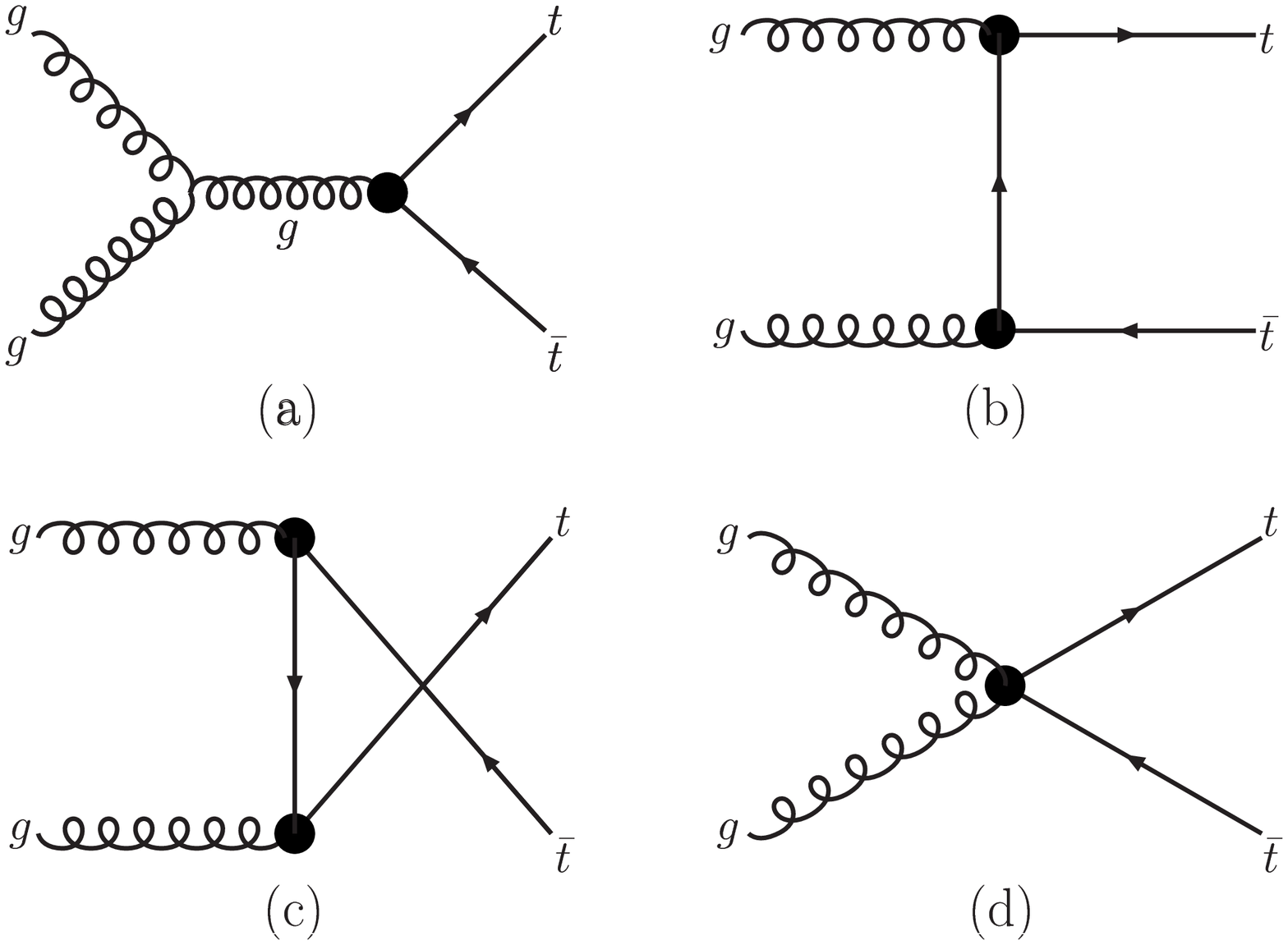}}
\caption{Feynman diagrams of $gg \to t\bar{t}$. The bullet $\bullet$ expresses
the vertex which includes the anomalous couplings.}\label{Feyngg}
\end{minipage}
\end{figure}
\end{center}

On the other hand, $g g\to t\bar{t}$ consists of four intermediate states (Fig.\ref{Feyngg} a,b,c,d),
and the corresponding amplitudes are
\begin{eqnarray}
&&{\cal M}_{gg}
={\cal M}_{gg}^{\rm a} + {\cal M}_{gg}^{\rm b} + {\cal M}_{gg}^{\rm c}
 + {\cal M}_{gg}^{\rm d},
\nonumber \\
&&\ \ \ {\cal M}_{gg}^{\rm a}
= -\frac{g_s^2}{2\hat{s}} \sum_a \bar{u}(\mib{p}_t) \lambda^a {\mit\Gamma}^\mu(q)
v(\mib{p}_{\bar{t}}) \nonumber \\
&&\phantom{{\cal M}_{gg}^1}\ \ \ \
\times if_{abc}[\:2q_{2\,\nu} \epsilon^\nu(\mib{q}_1) \epsilon_\mu(\mib{q}_2)
- 2q_{1\,\nu} \epsilon_\mu (\mib{q}_1) \epsilon^\nu(\mib{q}_2)
\nonumber \\
&&\phantom{{\cal M}_{gg}^1}\ \ \ \ \ \ \ \ \ \ \ \ \ \ \ \
+(q_1 -q_2)_\mu \epsilon_\nu (\mib{q}_1) \epsilon^\nu(\mib{q}_2)\:]
\\
&&\ \ \ {\cal M}_{gg}^{\rm b}
=\frac14 g_s^2 \,\bar{u}(\mib{p}_t) \lambda^b\lambda^c {\mit\Gamma}^\mu(q_1)
\frac1{m_t-\sla{k}_1} {\mit\Gamma}^\nu(q_2) v(\mib{p}_{\bar{t}}) \,
\epsilon_\mu (\mib{q}_1) \epsilon_\nu(\mib{q}_2)
\\
&&\ \ \ {\cal M}_{gg}^{\rm c}
=\frac14 g_s^2 \,\bar{u}(\mib{p}_t) \lambda^c\lambda^b {\mit\Gamma}^\mu(q_2)
\frac1{m_t-\sla{k}_2}
{\mit\Gamma}^\nu(q_1) v(\mib{p}_{\bar{t}}) \,
\epsilon_\nu (\mib{q}_1) \epsilon_\mu(\mib{q}_2)
\\
&&\ \ \ {\cal M}_{gg}^{\rm d}
=- g_s^2 \sum_a f_{abc} \bar{u}(\mib{p}_t) \lambda^a {\mit\Sigma}^{\mu\nu}
v(\mib{p}_{\bar{t}})\,
\epsilon_\mu (\mib{q}_1) \epsilon_\nu(\mib{q}_2).
\end{eqnarray}
Here $k_1 \equiv p_t -q_1$,\ $k_2 \equiv p_t -q_2$,\ $[a]$ and $[b,\ c]$ are the
color labels of the intermediate gluon and the incident gluons with momenta $q_1,\ q_2$,
$\epsilon(\mib{q}_{1,2})$ are the incident-gluon polarization vectors, and
\[
{\mit\Sigma}^{\mu\nu} \equiv \frac{\sigma^{\mu\nu}}{m_t} (d_V + id_A \gamma_5).
\]

\vskip 0.6cm

Based on these invariant amplitudes, the differential cross sections are calculated:
the one for $q\bar{q} \to t\bar{t}$ in the $q\bar{q}$-CM frame is
\begin{eqnarray}
&&\frac{\ \ d\sigma_{q\bar{q}}}{dE_t^* d\cos\theta_t^*}
= \frac{\beta_t^*}{16\pi \hat{s}}\delta(\sqrt{\hat{s}}-2E_t^*)
\Bigl(\frac13\Bigr)^2\sum_{\rm color}\Bigl(\frac12\Bigr)^2
\sum_{\rm spin}|{\cal M}_{q\bar{q}}|^2,~~~~~
\label{qqtt}
\end{eqnarray}
and the one for $gg \to t\bar{t}$ in the $gg$-CM frame is
\begin{eqnarray}
&&\frac{\ \ d\sigma_{gg}}{dE_t^* d\cos\theta_t^*}
= \frac{\beta_t^*}{16\pi \hat{s}}\delta(\sqrt{\hat{s}}-2E_t^*)
\Bigl(\frac18\Bigr)^2\sum_{\rm color}\Bigl(\frac12\Bigr)^2
\sum_{\rm spin}|{\cal M}_{gg}|^2,~~~~~
\label{ggtt}
\end{eqnarray}
where the asterisk was used to express that the quantities with it are those in the parton
CM frame, $\beta_t^* \equiv |\mib{p}_t^*|/E_t^* (= \sqrt{1-4m_t^2/\hat{s}})$ is the size of the
produced top-quark velocity in this frame, and we already performed the azimuthal-angle integration
since there is no non-trivial dependence on this angle included. After carrying out the color
summation, we use the
algebraic calculation system FORM \cite{FORM} to evaluate $|{\cal M}|^2$, and perform numerical
computations.

Concerning analytical expression of $\sum|{\cal M}|^2$, compact formulas are
found in \cite{Haberl:1995ek,Gluck:1977zm}, which lead to
\begin{eqnarray}
&&\!\!\!\!\!\!\!{\sum_{\rm color}\sum_{\rm spin}|{\cal M}_{q\bar{q}}|^2}
= 16 g_s^4\,\Bigl[\,1-2(v-z)-8(d_V-d_V^2+d_A^2) % \nonumber\\
% &&\phantom{\sum_{\rm color}\sum_{\rm spin}|{\cal M}_{gg}|^2= 16 g_s^2\,\Bigl[\,}
+8(d_V^2+d_A^2)v/z\,\Bigr],~~~~~ \label{Mqq}
\\
&&\!\!\!\!\!\!\!{\sum_{\rm color}\sum_{\rm spin}|{\cal M}_{gg}|^2}
= \frac{32}{3}g_s^4\,\Bigl[\,(4/v-9)\,[\,1-2v+4z(1-z/v)-8d_V(1-2d_V)\,] \nonumber\\
&&\!\!\!\!\!\!\!\!\!\!
\phantom{\sum_{\rm color}\sum_{\rm spin}|{\cal M}_{gg}|^2= \frac{32}{3}g_s^2\,\Bigl[\,}
+4(d_V^2+d_A^2)\,[\,14(1-4d_V)/z+(1+10d_V)/v\,] \nonumber\\
&&\!\!\!\!\!\!\!\!\!\!
\phantom{\sum_{\rm color}\sum_{\rm spin}|{\cal M}_{gg}|^2= \frac{32}{3}g_s^2\,\Bigl[\,}
-32(d_V^2+d_A^2)^2(1/z-1/v-4v/z^2)\,\Bigr], \label{Mgg}
\end{eqnarray}
where $z\equiv m_t^2/\hat{s}$, $v\equiv (\hat{t}-m_t^2)(m_t^2-\hat{s}-\hat{t})/\hat{s}^2$,
$\hat{t}\equiv (q_1-p_t)^2$, we have re-expressed their anomalous-coupling parameters
in our notation, and the overall 4-momentum conservation has been taken into account.
We confirmed that our FORM results are in complete agreement with them.

When we derive hadron cross sections from parton cross sections, we first need to
connect parton cross sections in the parton-CM frame and hadron-CM frame. Among the quantities
we are going to compute here, the total cross section, the top-quark $p_{\rm T}$ distribution,
and the $t\bar{t}$ invariant-mass distributions are all Lorentz-transformation
invariant,\footnote{Concerning the $p_{\rm T}$ distribution, note that we do not take
    into account the parton transverse momenta.}\ 
therefore we do not have to worry about the difference between these two frames. 
In case of the $p_{\rm T}$ distribution, we have
\begin{eqnarray}
&&\frac{d}{dp_{\rm T}}\sigma_{q\bar{q},gg}
= \frac{d}{dp^*_{\rm T}}\sigma_{q\bar{q},gg}
= \int^{c_{\rm max}^*}_{c_{\rm min}^*} d\cos\theta_t^*
\frac{\ \:\:d\sigma_{q\bar{q},gg}}{dp^*_{\rm T} d\cos\theta_t^*}  \nonumber \\
&&\phantom{\frac{d}{dp_{\rm T}}\sigma_{q\bar{q},gg}}
= \int^{c_{\rm max}^*}_{c_{\rm min}^*} d\cos\theta_t^*
\frac{|\mib{p}_t^*|}{E_t^* \sin\theta_t^*}
\frac{\ \:\:d\sigma_{q\bar{q},gg}}{dE_t^* d\cos\theta_t^*},
\end{eqnarray}
where the quantities without $*$ are those in the hadron-CM frame, we have chosen
the $z$ axis in the direction of $\mib{q}_1$ so that $p_{\rm T}^*(=p_{\rm T}) = 
|\mib{p}_t^*|\sin\theta_t^*$ and
\begin{equation}
c_{\rm max}^* = -c_{\rm min}^* = \sqrt{1-4p_{\rm T}^{*2}/(\beta_t^*\sqrt{\hat{s}})^2},
\end{equation}
and for the $t\bar{t}$ invariant-mass distribution,
\begin{equation}
\frac{d}{d\mu_{t\bar{t}}}\sigma_{q\bar{q},gg}
= \frac{d}{d\mu^*_{t\bar{t}}}\sigma_{q\bar{q},gg}
= \frac12 \frac{d}{dE^*_t}\sigma_{q\bar{q},gg}
= \frac12 \int^{+1}_{-1} d\cos\theta_t^*
\frac{d\sigma_{q\bar{q},gg}}{dE_t^* d\cos\theta_t^*},
\end{equation}
where $\mu_{t\bar{t}}=\mu_{t\bar{t}}^* = \sqrt{\hat{s}} = 2E_t^*$.
On the other hand,
we need the appropriate Jacobian connecting the two frames when the angular distribution is
considered as follows: The top energy and scattering angle in the parton-CM frame are expressed in
terms of those in the hadron-CM frame as
\begin{eqnarray}
&&E_t^*=(E_t - \beta |\mib{p}_t|\cos\theta_t)/\sqrt{1-\beta^2}, \\
&&\cos\theta_t^*=(|\mib{p}_t|\cos\theta_t-\beta E_t)/(|\mib{p}_t^*|\sqrt{1-\beta^2}),
\label{costhetat}
\end{eqnarray}
where $\beta$ is the Lorentz-transformation boost factor connecting the two frames, and we
used $|\mib{p}_t^*|\,(=\sqrt{E_t^{*2}-m_t^2})$ in the denominator on the right-hand side of
eq.(\ref{costhetat}) to make the formula compact. These relations lead to the Jacobian
\begin{equation}
\partial(E_t^*,\,\cos\theta_t^*)/\partial(E_t,\,\cos\theta_t)
=|\mib{p}_t|/|\mib{p}_t^*|
\end{equation}
and the cross-section relation
\begin{equation}
\frac{\ \ \ d\sigma_{q\bar{q},gg}}{dE_t d\cos\theta_t}
=\frac{|\mib{p}_t|}{|\mib{p}_t^*|}
\frac{\ \ \ d\sigma_{q\bar{q},gg}}{dE_t^* d\cos\theta_t^*}\,.
\end{equation}

Then the hadron cross sections are obtained by integrating the product of
{\it the parton distribution functions} and {\it the parton cross sections in
the hadron-CM frame} on the momentum
fractions $x_1$ and $x_2$ carried by the partons. Let us explicitly show the result
of the above $E_t$ and $\theta_t$ double distribution, since the other quantities
are easier to handle: 
\begin{equation}
\frac{\ \ \ d\sigma_{p\bar{p}/pp}}{dE_t d\cos\theta_t}
= \sum_{a,b} \int^1_{4m_t^2/s}\! \!\!\!dx_1 \int^1_{4m_t^2/(x_1 s)} \!\!\!\!\!\!\!\!dx_2
\:N_a(x_1) N_b(x_2)
\frac{|\mib{p}_t|}{|\mib{p}_t^*|}
\frac{\ d\sigma_{ab}}{dE_t^* d\cos\theta_t^*},
\end{equation}
where $N_{a,b}(x)$ are the parton distribution functions of parton $a$ and $b$
($a,b=u,\bar{u}$, $d,\bar{d}$, $s,\bar{s}$, $c,\bar{c}$, $b,\bar{b}$ and $g$).
Thanks to
the energy-conservation delta function in Eqs.(\ref{qqtt}) and (\ref{ggtt}),
we can immediately perform $x_2$ integration and get to
\begin{eqnarray}
&&\frac{\ \ \ d\sigma_{p\bar{p}/pp}}{dE_t d\cos\theta_t}
= \sum_{a,b} \int^1_{x_{1\,{\rm min}}} \!\!\!dx_1 N_a(x_1) N_b(x_2)
 \frac{x_2 \beta_t \displaystyle\sqrt{1-\beta^2}}{(1+\beta)(1-\beta_t\cos\theta_t)}  \nonumber \\
&&
\phantom{\frac{d\sigma}{dE_t d\cos\theta_t}
= \sum_{a,b} \int^1_{x_{1\,{\rm min}}}}
\times \frac1{8\pi \hat{s}\sqrt{\hat{s}}} \Bigl(\frac1{f_c}\Bigr)^2\sum_{\rm color}
\Bigl(\frac1{f_s}\Bigr)^2 \sum_{\rm color}
|{\cal M}_{ab}|^2,~~
\label{Dsigma2}
\end{eqnarray}
where $\beta_t \equiv |\mib{p}_t|/E_t$, $\beta = (x_1-x_2)/(x_1+x_2)$, $\hat{s}$
is related to $s$ defined via the
hadron momenta $p_{1,2}$ ($s \equiv (p_1 + p_2)^2$) as $\hat{s}=x_1 x_2 s$, $f_c$ and $f_s$
are the color
and spin degrees of freedom of the incident partons respectively, and $x_2$ is now
given as
\begin{equation}
x_2=\frac{x_1 E_t (1-\beta_t\cos\theta_t)}{x_1\sqrt{s}-E_t(1+\beta_t\cos\theta_t)}.
\end{equation}
Since $x_1$ and $x_2$ must satisfy $4m_t^2/(x_1 s) \leq x_2 \leq 1$, we have
\begin{equation}
x_{1\,{\rm min}}
=\frac{E_t (1+\beta_t\cos\theta_t)}{\sqrt{s}-E_t(1-\beta_t\cos\theta_t)}.
\end{equation}
The top angular distribution is obtained by integrating eq.(\ref{Dsigma2}) on $E_t$
over
\[
m_t \leq E_t \leq \sqrt{s}/2.
\]

% 333333333333333333333333333333333333333333333333333333333333
\sec{Analyses}

We are now ready to perform numerical computations. We first compare the total
cross section of $p\bar{p}\to t\bar{t}X$ at Tevatron energy with CDF/D0 data
to get improved constraints on $d_{V,A}$, then compute the total cross section,
the top angular distribution, the top $p_{\rm T}$ distribution, and the $t\bar{t}$
invariant-mass distribution of $pp \to t\bar{t}X$ at LHC energy.
Those top cross sections are not a quantity which directly shows the $C\!P$ nature
of the interactions, and therefore depend both on $d_{V,A}$. This may seem
inefficient but we could thereby get useful information of both parameters
at the same time.

\subsec{Analysis of Tevatron data}

The latest data of $t\bar{t}$ pair productions at Tevatron for $\sqrt{s}=1.96$ TeV
are \cite{Teva-data}
\begin{eqnarray}
&&\!\!\!\!\!\!\!\!\!
\sigma_{\rm exp}
= 7.02 \pm 0.63\ {\rm pb}\: \ \ ({\rm CDF}:\:m_t=175\:{\rm GeV}) \\
&&\!\!\!\!\!\!\!\!\!
\phantom{\sigma_{\rm exp}}
= 8.18^{\ +\ 0.98}_{\ -\ 0.87}\ {\rm pb}\ \ \ \ ({\rm D0}:\:m_t=170\:{\rm GeV}).
\end{eqnarray}
We could decrease the uncertainty if we combined them according to the standard formula
of statistics, and indeed such averaging is often seen in many papers. We, however, 
would rather stay conservative and do not follow this way because it is not easy to
treat (average) properly systematic errors in different detectors. In addition,
different values are used for $m_t$ in their analyses, which also makes the averaging
difficult. 
At any rate, the uncertainties of CDF and D0 data were $+3.6/-2.4$ pb and $\pm 2.2$ pb
respectively when Haberl et al. performed the analysis \cite{Haberl:1995ek}, which tells us
that it is truly the time to revisit this analysis. 

On the other hand, the total cross section in the framework of QCD with higher order
corrections has been studied in detail in \cite{Kidonakis:2008mu} (see also \cite{Moch:2008qy}).
We take the results using the latest set of parton-distribution functions ``CTEQ6.6M"
(NNLO approximation) \cite{Nadolsky:2008zw}
\begin{eqnarray}
&&\!\!\!\!\!\!\!\!\!
\sigma_{\rm QCD}
= 6.73^{\ +\ 0.51}_{\ -\ 0.46}\ {\rm pb}\: \ \ (m_t=175\:{\rm GeV}) \\
&&\!\!\!\!\!\!\!\!\!
\phantom{\sigma_{\rm QCD}}
= 7.87^{\ +\ 0.60}_{\ -\ 0.55}\ {\rm pb}\: \ \ (m_t=170\:{\rm GeV}),
\end{eqnarray}
and combine these theoretical errors with the above experimental errors as
\begin{eqnarray}
&&\!\!\!\!\!\!\!\!\!
\sigma_{\rm exp}
= 7.02^{\ +\ 0.81}_{\ -\ 0.78}\ {\rm pb}\: \ \ ({\rm CDF}:\:m_t=175\:{\rm GeV}) \\
&&\!\!\!\!\!\!\!\!\!
\phantom{\sigma_{\rm exp}}
= 8.18^{\ +\ 1.15}_{\ -\ 1.03}\ {\rm pb}\: \ \ ({\rm D0}:\:m_t=170\:{\rm GeV}).
\label{sigmadata}
\end{eqnarray}

Comparing them with our calculations $\sigma(d_V, d_A)$, which is the sum of the central
values of the above $\sigma_{\rm QCD}$ and the non-SM part of our cross sections at the
lowest order of perturbation, we find that $d_{V,A}$ are restricted as
\begin{eqnarray}
&&-0.01\ \rs\ d_V\ \rs\:+0.01 \ \ \ \  \ +0.38\ \rs\ d_V\ \rs\:+0.41
\end{eqnarray}
when we put $d_A=0$. Similarly we have
\begin{equation}
|d_A|\ \rs\:+0.12
\end{equation}
when we put $d_V=0$. Here, since $\sigma(d_V, d_A)$ depends on not $d_A$ but $d_A^2$ as is
known from eqs.(\ref{Mqq}) and (\ref{Mgg}), we only
get constraints on $|d_A|$. Finally, when we keep both $d_{V,A}$ non-zero, these two
parameters produce corrections which tend to cancel each other unless $|d_V|$ is
not that sizable, and consequently rather large $d_{V,A}$ are allowed:
\begin{equation}
d_V \simeq +0.2\ \ \ {\rm and}\ \ \ |d_A| \simeq +0.3.
\end{equation}

\begin{figure}[htbp]
\begin{minipage}{14.8cm}
\begin{center}
% \hspace*{1.5cm}
\psfrag{dv}{\begin{large}\hspace*{-0.0cm}$d_V$\end{large}}
\psfrag{da}{\begin{large}\hspace*{-0.0cm}$d_A$\end{large}}
{\includegraphics[width=13.95cm]{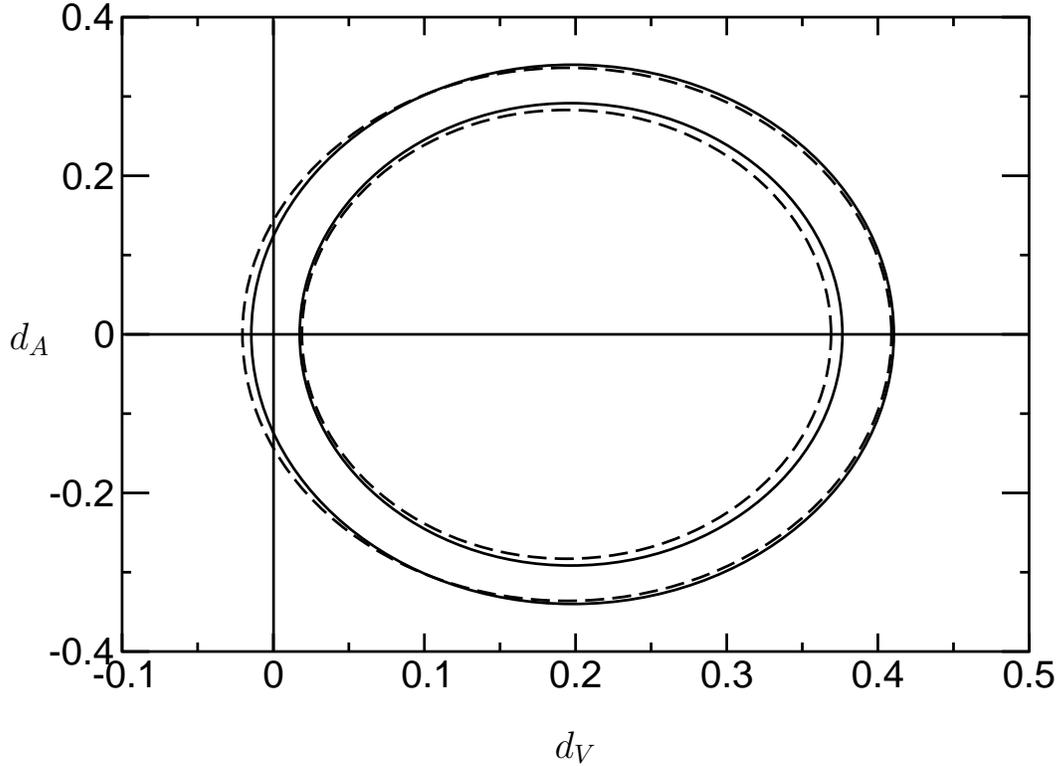}}
% {\includegraphics[width=8cm]{contor.eps}}
\caption{Experimentally allowed region for $d_{V,A}$. The region between two solid/dashed curves is
from CDF/D0 data.}\label{contor1}
\end{center}
\end{minipage}
\end{figure}

We show the experimentally allowed $d_{V,A}$ region in Fig.\ref{contor1}. We find that there still remains
some area for these anomalous-coupling parameters, though the standard-model (QCD) prediction,
i.e., $d_{V,A}=0$ is consistent with the data, too.

\vskip 3cm % \vfill

% \end{center}

\newpage

\subsec{LHC I: Total cross sections}

Let us compute the total cross section and the differential distributions
of $pp \to t\bar{t}X$ at LHC energy ($\sqrt{s}=$ 10 and 14 TeV) for

\begin{center}
$(d_V,\:d_A)=$\ \ \
(a)\ $(-0.01,\:0)$,\ \ \
(b)\ $(0.41,\:0)$,\ \ \
(c)\ $(0,\:0.12)$,\ \ \
(d)\ $(0.2,\:0.3)$
\end{center}
as typical examples. Concerning the top-quark mass, we use the present world average
$m_t=172$ GeV \cite{:2008vn}.

First, the total cross sections of top pair productions are:

$\sqrt{s}=$ 10 TeV
\begin{equation}
\begin{array}{llrl}
{\rm (a)}\ \ d_V = -0.01,\ d_A=0~~~~~~& \sigma =& 447 &{\rm pb}\\
{\rm (b)}\ \ d_V = 0.41, \ d_A=0      & \sigma =& 1240 &{\rm pb}\\
{\rm (c)}\ \ d_V = 0 ,\ d_A= 0.12     & \sigma =& 637 &{\rm pb}\\
{\rm (d)}\ \ d_V = 0.2,\ d_A = 0.3    & \sigma =& 1835 &{\rm pb} \\
\end{array}
\end{equation}

$\sqrt{s}=$ 14 TeV
\begin{equation}
\begin{array}{llrl}
{\rm (a)}\ \ d_V = -0.01,\ d_A=0~~~~~~& \sigma =& 991 &{\rm pb}\\
{\rm (b)}\ \ d_V = 0.41, \ d_A=0      & \sigma =& 3479 &{\rm pb}\\
{\rm (c)}\ \ d_V = 0 ,\    d_A= 0.12  & \sigma =& 1458   &{\rm pb}\\
{\rm (d)}\ \ d_V = 0.2,\   d_A = 0.3  & \sigma =& 4744   &{\rm pb}
\end{array}
\end{equation}
They are much larger than the latest QCD predictions \cite{Kidonakis:2008mu}
\[
\begin{array}{l}
\sigma_{\rm QCD}(\sqrt{s}=10\ {\rm TeV})= 415^{\ +\ 34}_{\ -\ 29}\ {\rm pb}, \\
\sigma_{\rm QCD}(\sqrt{s}=14\ {\rm TeV})= 919^{\ +\ 76}_{\ -\ 55}\ {\rm pb}. \\
\end{array}
\]
In particular, the result with $(d_V,\:d_A)=(0.2,\:0.3)$ is several times larger than
$\sigma_{\rm QCD}$, which means that we might encounter a surprising observation at LHC.

\begin{figure}[htbp]
\begin{minipage}{14.8cm}
\begin{center}
% \hspace*{1.5cm}
\psfrag{dv}{\begin{large}\hspace*{-0.0cm}$d_V$\end{large}}
\psfrag{da}{\begin{large}\hspace*{-0.0cm}$d_A$\end{large}}
{\includegraphics[width=10.cm]{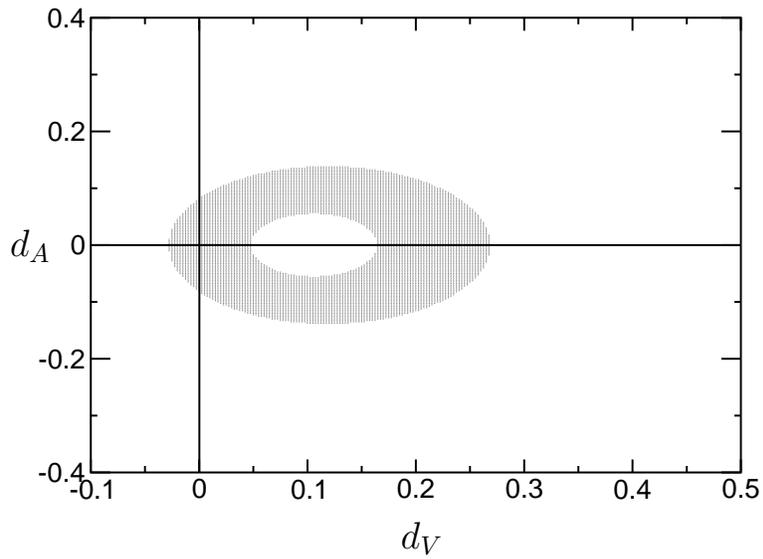}}
\caption{Allowed region for $d_{V,A}$ which LHC ($\sqrt{s}=$10 TeV) might give us.}\label{contor2}
\end{center}
\end{minipage}
\end{figure}

\begin{figure}[htbp]
\begin{minipage}{14.8cm}
\begin{center}
% \hspace*{1.5cm}
\psfrag{dv}{\begin{large}\hspace*{-0.0cm}$d_V$\end{large}}
\psfrag{da}{\begin{large}\hspace*{-0.0cm}$d_A$\end{large}}
{\includegraphics[width=10.cm]{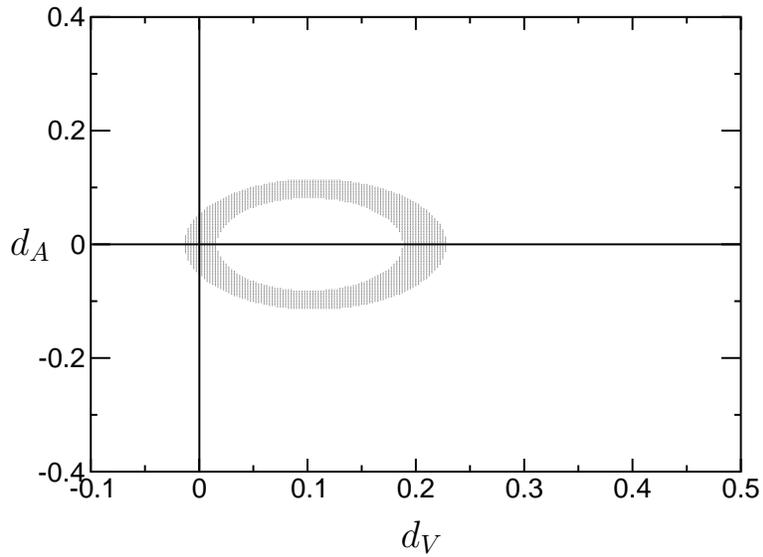}}
\caption{Allowed region for $d_{V,A}$ which LHC ($\sqrt{s}=$14 TeV) might give us.}\label{contor3}
\end{center}
\end{minipage}
\end{figure}

It is indeed remarkable that the present Tevatron data still allow such a huge
cross section at LHC, but this also indicates that coming measurements at LHC 
might give us a much stronger constraint on $d_{V,A}$. In order to see this
possibility clearly, we assume that we have
\[
\begin{array}{l}
\sigma (\sqrt{s}=10\ {\rm TeV})= 415 \pm 100 \ {\rm pb}, \\
\sigma (\sqrt{s}=14\ {\rm TeV})= 919 \pm 100 \ {\rm pb}  \\
\end{array}
\]
at LHC (including possible theoretical uncertainties), and draw figures similar to
Fig.\ref{contor1} in Figs.\ref{contor2} \& \ref{contor3}.
They show that LHC will actually give a very good opportunity to perform precise analyses of
top-gluon couplings.

\newpage % \vspace{0.95cm}

% \end{center}

\vspace{1.cm}

% \newpage % 
\vspace{0.8cm}

\begin{figure}[htbp]
\begin{minipage}{14.8cm}
\begin{center}
% \hspace*{1.5cm}
% \psfrag{dv}{\begin{large}\hspace*{-0.0cm}$d_V$\end{large}}
% \psfrag{da}{\begin{large}\hspace*{-0.0cm}$d_A$\end{large}}
{\includegraphics[width=14cm]{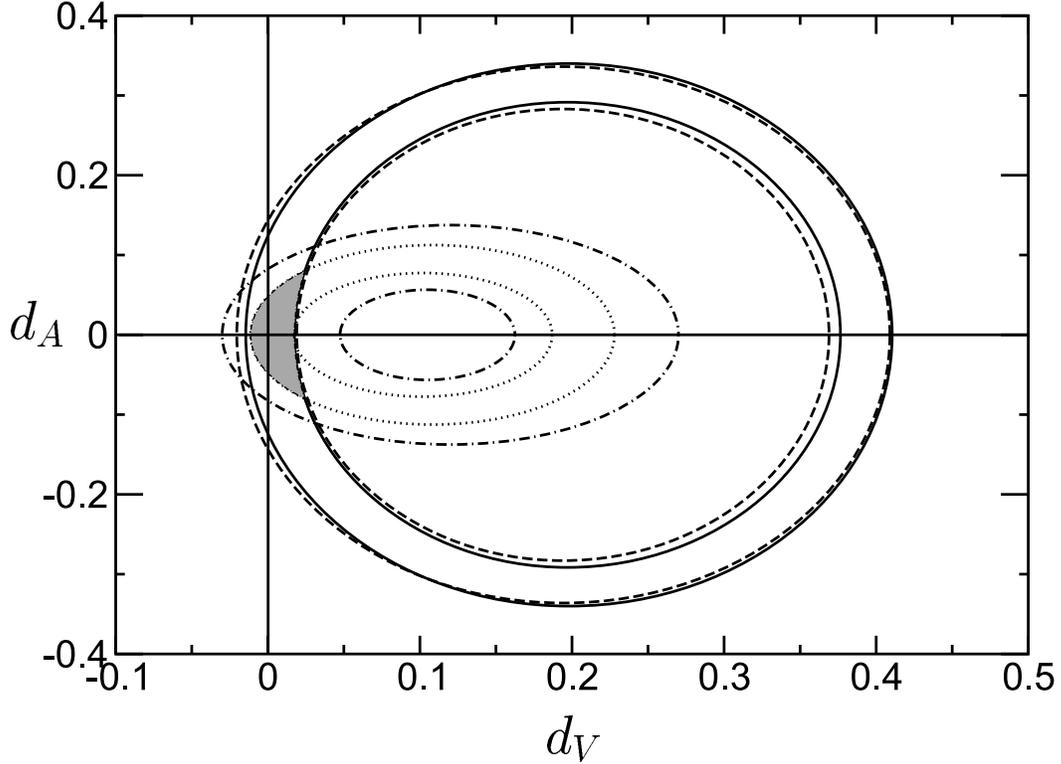}}
\caption{The $d_{V,A}$ region allowed by Tevatron and {\it assumed} LHC data
(the shaded part).}\label{allow}
\end{center}
\end{minipage}
\end{figure}

In this {\it virtual} analysis, however, one may expect that $d_V \simeq d_A \simeq 0$,
i.e., an area around the QCD prediction is chosen as the best and unique solution since we used the very QCD result
for the central value of the {\it assumed} data, but what the two figures show us seems to be against
this expectation. This is due to the cancellation between the $d_V$ and $d_A$ terms as mentioned in
the previous subsection. Then, is it
impossible to single out QCD even if we got much more precise data as long as we rely on the total
cross section alone? Fortunately it is not right: superposing the constraints from Tevatron and LHC,
we find that only a small region around $d_V=d_A =0$ would survive as in Fig.\ref{allow} if
the above {\it assumed} LHC data were true.

\vspace{1.5cm}

% \newpage

\vspace{1cm}

\subsec{LHC I\hspace{-0.06cm}I: Differential distributions}

\begin{figure}[h]
\begin{minipage}{14.8cm}
\begin{center}
\psfrag{dsig}{\begin{Large}\hspace*{-1.7cm}
$\frac1{\sigma_0}\frac{d\sigma}{d\cos\theta_t}$\end{Large}}
\psfrag{cs}{\begin{large}\hspace*{-0.0cm}$\cos\theta_t$\end{large}}
\psfrag{c1}{\begin{small}QCD\end{small}}
\psfrag{c2}{\begin{small}$d_V=-0.01,\:d_A=0$\end{small}}
\psfrag{c3}{\begin{small}$d_V=0.41,\:d_A=0$\end{small}}
\psfrag{c4}{\begin{small}$d_V=0,\:d_A=0.12$\end{small}}
\psfrag{c5}{\begin{small}$d_V=0.2,\:d_A=0.3$\end{small}}
{\includegraphics[width=8.3cm]{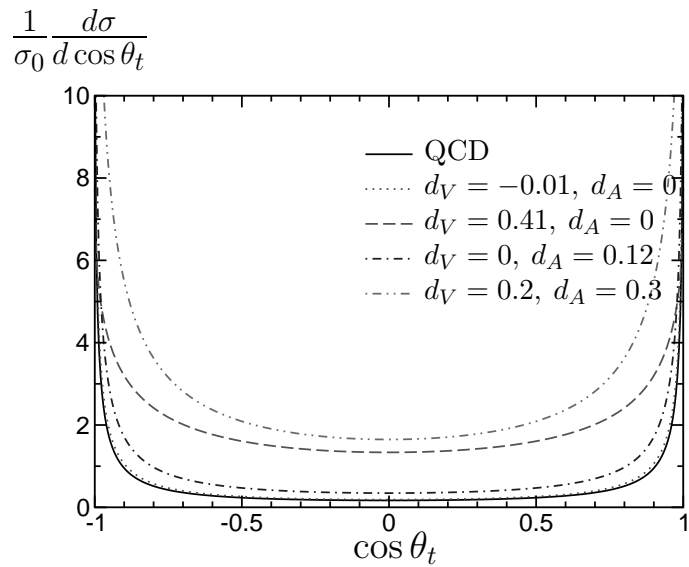}}
\caption{The top angular distribution normalized
by $\sigma_0$: LHC energy $\sqrt{s}=10$ TeV}\label{Dis10}
\end{center}
\end{minipage}
\end{figure}

\begin{figure}[h]
\begin{minipage}{14.8cm}
\begin{center}
\psfrag{dsig}{\begin{Large}\hspace*{-1.7cm}
$\frac1{\sigma_0}\frac{d\sigma}{d\cos\theta_t}$\end{Large}}
\psfrag{cs}{\begin{large}\hspace*{-0.0cm}$\cos\theta_t$\end{large}}
\psfrag{c1}{\begin{small}QCD\end{small}}
\psfrag{c2}{\begin{small}$d_V=-0.01,\:d_A=0$\end{small}}
\psfrag{c3}{\begin{small}$d_V=0.41,\:d_A=0$\end{small}}
\psfrag{c4}{\begin{small}$d_V=0,\:d_A=0.12$\end{small}}
\psfrag{c5}{\begin{small}$d_V=0.2,\:d_A=0.3$\end{small}}
{\includegraphics[width=8.3cm]{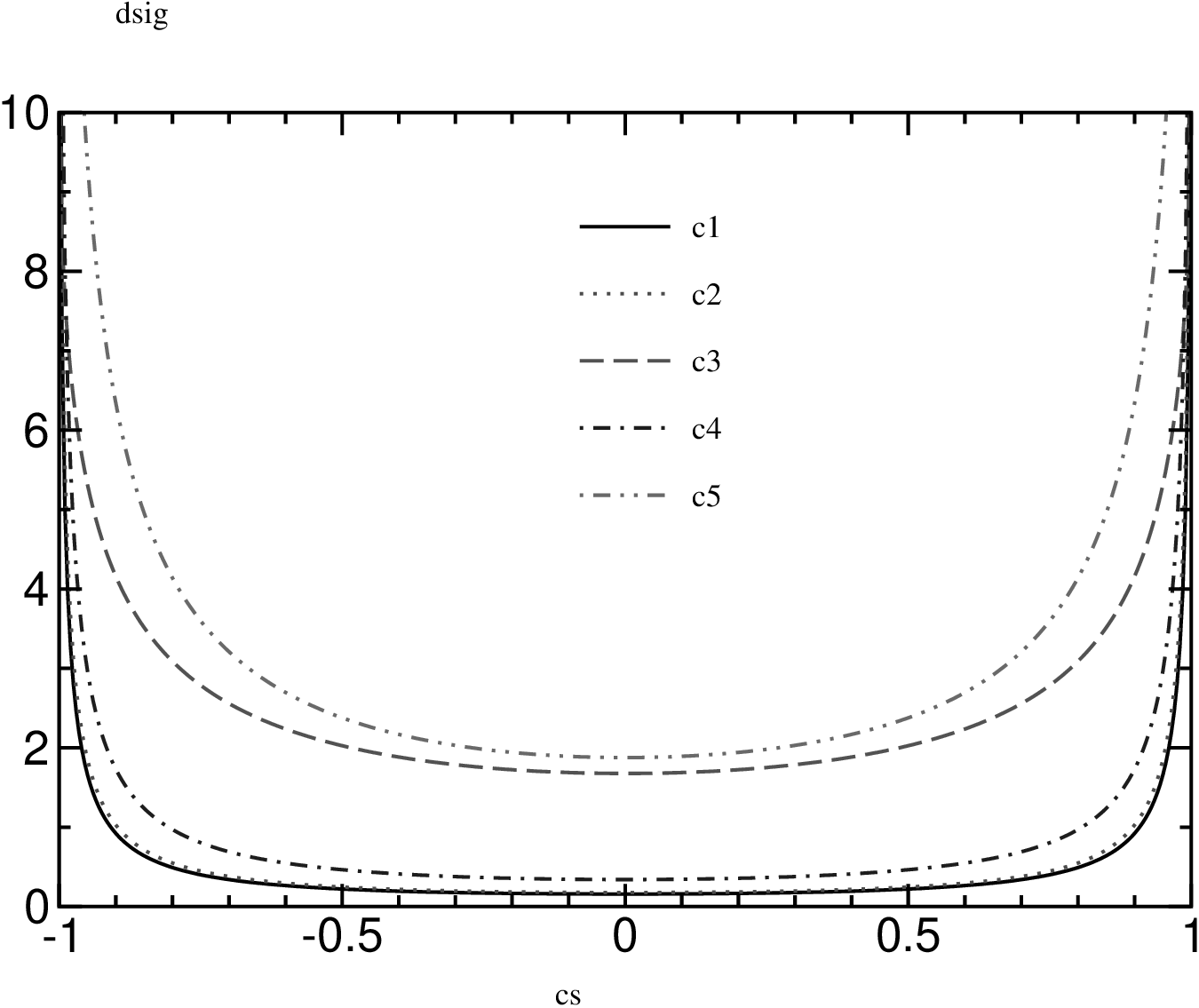}}
\caption{The top angular distribution normalized
by $\sigma_0$: LHC energy $\sqrt{s}=14$ TeV}\label{Dis14}
\end{center}
\end{minipage}
\end{figure}

Next, we give the top angular distributions in $pp \to t\bar{t}X$ normalized by
$\sigma_0 = \sigma(d_V=d_A=0)$, i.e., $\sigma_0^{-1}d\sigma/d\cos\theta_t$
in Figs.\ref{Dis10} and \ref{Dis14} for $\sqrt{s}=$ 10 TeV and 14 TeV, where both
$d\sigma$ and $\sigma_0$ are the tree-level quantities (concerning this approximation,
see the later comments). As was just
shown, the size of the cross section becomes larger when $d_{V,A} \neq 0$, so
the corresponding distributions normalized by $\sigma_0$ also exceed the QCD
result (solid curve), and moreover their shapes are different from the QCD distribution.

\vfill % \vspace{0.3cm} % \newpage

\vspace{0.8cm} % \newpage

% \end{center}

\newpage % \vspace{0.8cm} 

\begin{figure}[h]
\begin{minipage}{14.8cm}
\begin{center}
\psfrag{dsig}{\begin{Large}\hspace*{-1.7cm}
$\frac1{\sigma_0}\frac{d{\mit\Delta}\sigma}{d\cos\theta_t}$\end{Large}}
\psfrag{cs}{\begin{large}\hspace*{-0.0cm}$\cos\theta_t$\end{large}}
\psfrag{c2}{\begin{small}$d_V=-0.01,\:d_A=0$\end{small}}
\psfrag{c3}{\begin{small}$d_V=0.41,\:d_A=0$\end{small}}
\psfrag{c4}{\begin{small}$d_V=0,\:d_A=0.12$\end{small}}
\psfrag{c5}{\begin{small}$d_V=0.2,\:d_A=0.3$\end{small}}
{\includegraphics[width=8.3cm]{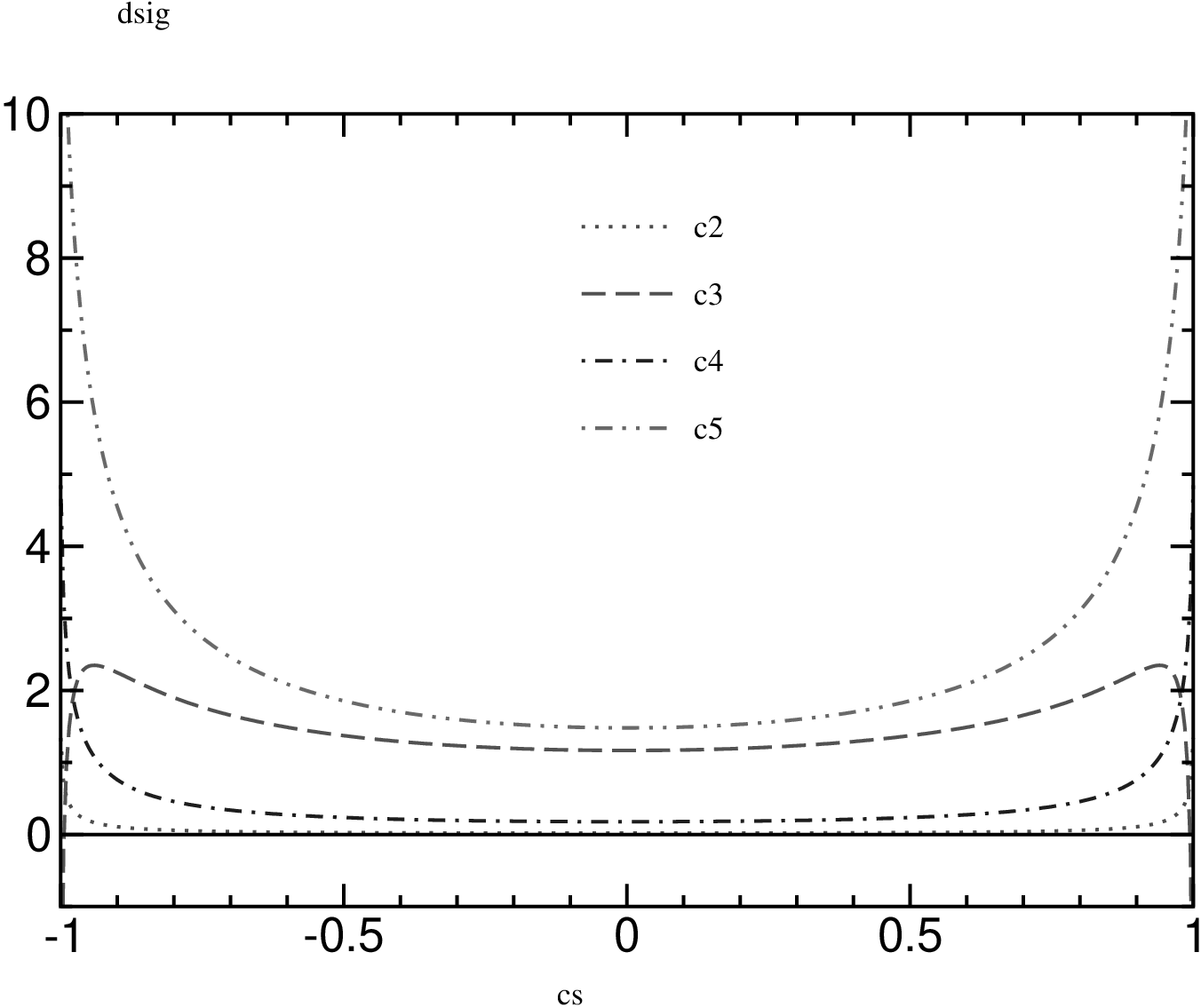}}
\caption{Nonstandard effects in the top angular distribution normalized
by $\sigma_0$: LHC energy $\sqrt{s}=10$ TeV}\label{Deldis10}
\end{center}
\end{minipage}
\end{figure}

\vspace{0.8cm} % \newpage

\begin{figure}[h]
\begin{minipage}{14.8cm}
\begin{center}
\psfrag{dsig}{\begin{Large}\hspace*{-1.7cm}
$\frac1{\sigma_0}\frac{d{\mit\Delta}\sigma}{d\cos\theta_t}$\end{Large}}
\psfrag{cs}{\begin{large}\hspace*{-0.0cm}$\cos\theta_t$\end{large}}
\psfrag{c2}{\begin{small}$d_V=-0.01,\:d_A=0$\end{small}}
\psfrag{c3}{\begin{small}$d_V=0.41,\:d_A=0$\end{small}}
\psfrag{c4}{\begin{small}$d_V=0,\:d_A=0.12$\end{small}}
\psfrag{c5}{\begin{small}$d_V=0.2,\:d_A=0.3$\end{small}}
{\includegraphics[width=8.3cm]{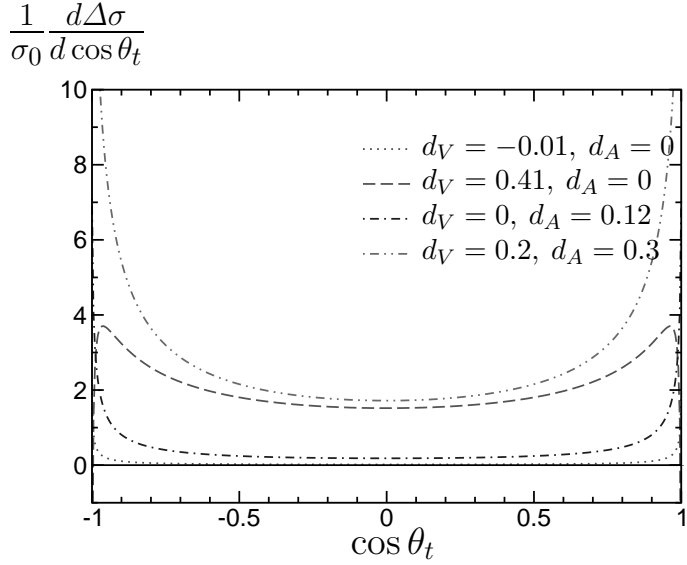}}
\caption{Nonstandard effects in the top angular distribution normalized
by $\sigma_0$: LHC energy $\sqrt{s}=14$ TeV}\label{Deldis14}
\end{center}
\end{minipage}
\end{figure}

\vspace{0.95cm}

Here some comments are necessary about the QCD radiative corrections. We calculated these
distributions at the lowest order in perturbation, assuming that most part of the corrections
to the standard-model cross sections $\sigma_0$ is canceled through the normalization between
the numerator and denominator, like the authors of \cite{Haberl:1995ek} did. According to
\cite{Beenakker:1990maa}, however, we should not rely on this approximation too much.
Therefore, we also show in Figs.\ref{Deldis10} \& \ref{Deldis14} the pure
nonstandard contribution $d{\mit\Delta}\sigma(d_V,d_A)\equiv d\sigma-d\sigma_0$, where
we normalize them by the same lowest-order $\sigma_0$ so that we can directly compare
Figs.\ref{Dis10} \& \ref{Dis14} and Figs.\ref{Deldis10} \& \ref{Deldis14}.
We find that all the curves there are similar to those in the previous figures except that
the curve for $d_V=0.41$ and $d_A=0$ (the dashed curve) behaves differently when $|\cos\theta_t|$
gets close to 1.

\begin{figure}[h]
\begin{minipage}{14.8cm}
\begin{center}
\psfrag{dsig}{\begin{Large}\hspace*{-0.8cm}
$\frac1{\sigma_0}\frac{d\sigma}{dp_{\rm T}}$\end{Large}}
\psfrag{c1}{\begin{small}QCD\end{small}}
\psfrag{c2}{\begin{small}$d_V=-0.01,\:d_A=0$\end{small}}
\psfrag{c3}{\begin{small}$d_V=0.41,\:d_A=0$\end{small}}
\psfrag{c4}{\begin{small}$d_V=0,\:d_A=0.12$\end{small}}
\psfrag{c5}{\begin{small}$d_V=0.2,\:d_A=0.3$\end{small}}
\psfrag{pt}{\begin{Large}$p_{\rm T}$\end{Large} (GeV)}
% \psfrag{gev}{\begin{large}GeV\end{large}}
{\includegraphics[width=12cm]{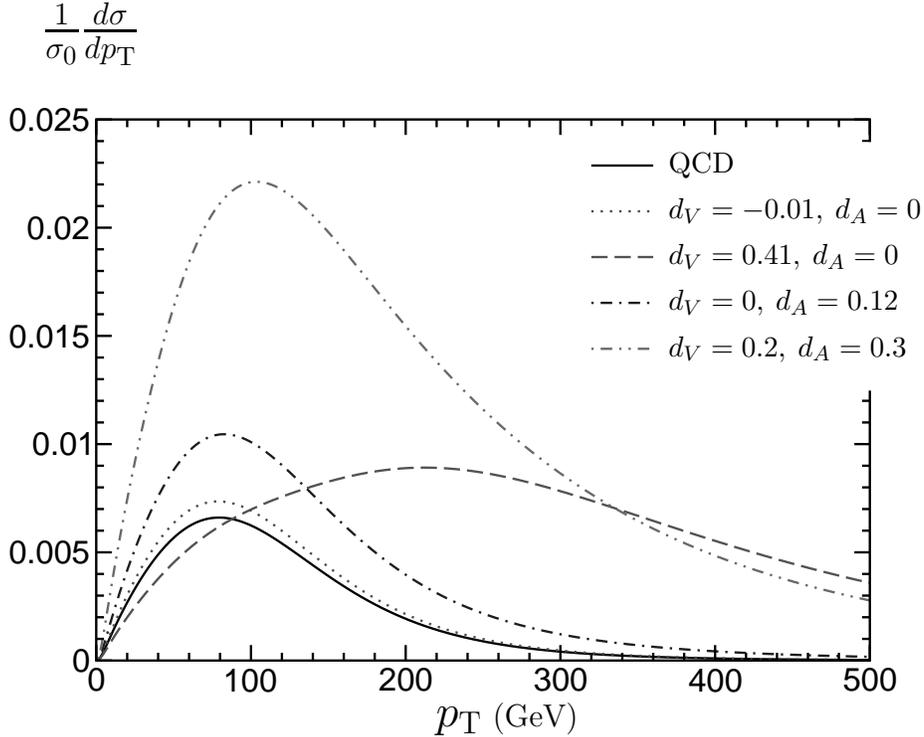}}
\caption{The top $p_{\rm T}$ distribution normalized
by $\sigma_0$: LHC energy $\sqrt{s}=10$ TeV}\label{Dispt10}
\end{center}
\end{minipage}
\end{figure}

\begin{figure}[h]
\begin{minipage}{14.8cm}
\begin{center}
\psfrag{dsig}{\begin{Large}\hspace*{-0.8cm}
$\frac1{\sigma_0}\frac{d\sigma}{dp_{\rm T}}$\end{Large}}
\psfrag{c1}{\begin{small}QCD\end{small}}
\psfrag{c2}{\begin{small}$d_V=-0.01,\:d_A=0$\end{small}}
\psfrag{c3}{\begin{small}$d_V=0.41,\:d_A=0$\end{small}}
\psfrag{c4}{\begin{small}$d_V=0,\:d_A=0.12$\end{small}}
\psfrag{c5}{\begin{small}$d_V=0.2,\:d_A=0.3$\end{small}}
\psfrag{pt}{\begin{Large}$p_{\rm T}$\end{Large} (GeV)}
% \psfrag{gev}{\begin{large}GeV\end{large}}
{\includegraphics[width=12cm]{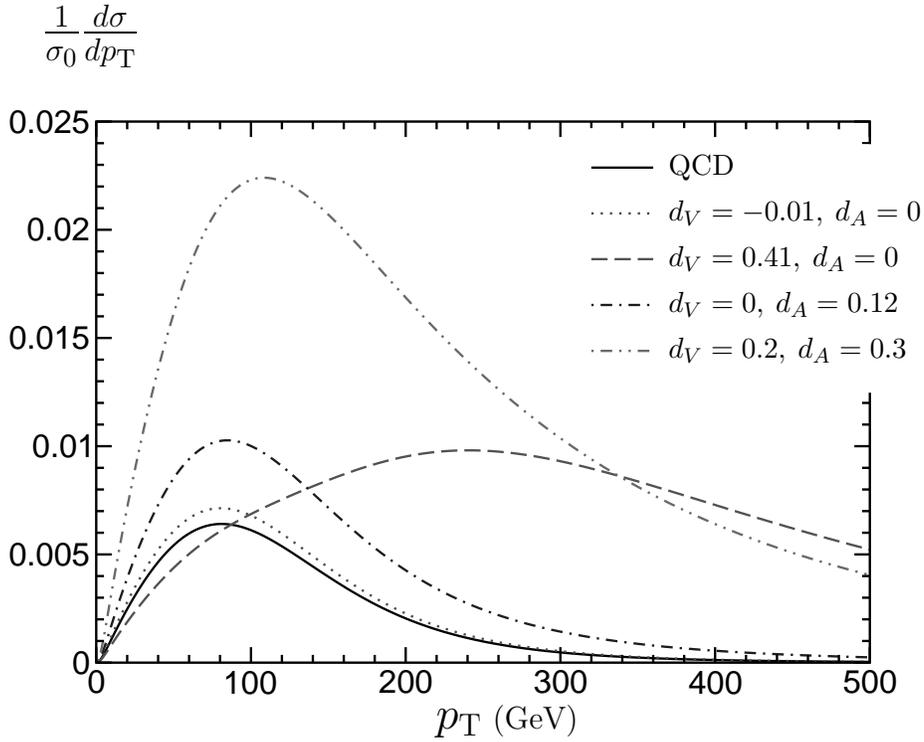}}
\caption{The top $p_{\rm T}$ distribution normalized
by $\sigma_0$: LHC energy $\sqrt{s}=14$ TeV}\label{Dispt14}
\end{center}
\end{minipage}
\end{figure}

In the same way, let us show the top $p_{\rm T}$ distributions $
\sigma_0^{-1}d\sigma/dp_{\rm T}$
in Figs.\ref{Dispt10} \& \ref{Dispt14}. There we give the whole distributions alone,
not the pure nonstandard contribution to them, since the above figures on
the angular distributions told us that the difference is not that significant.
We find that the shapes of four curves look rather alike, but the one for $(d_V,\,d_A)=(0.41,\,0)$
behaves differently and therefore apparently distinguishable from the others. 
In contrast to it, the difference
between the QCD curve and the one for $(d_V,\,d_A)=(-0.01,\,0)$ is so small that it will
be hard to draw meaningful information. 

\vfill % \vspace{0.3cm}

% \vspace{0.5cm} % 
\newpage

\vskip 0.4cm

\begin{figure}[h]
\begin{minipage}{14.8cm}
\begin{center}
\psfrag{dsig}{\begin{Large}\hspace*{-0.8cm}
$\frac1{\sigma_0}\frac{d\sigma}{d\mu_{t\bar{t}}}$\end{Large}}
\psfrag{cs}{\begin{large}\hspace*{-0.0cm}$\cos\theta_t$\end{large}}
\psfrag{c1}{\begin{small}QCD\end{small}}
\psfrag{c2}{\begin{small}$d_V=-0.01,\:d_A=0$\end{small}}
\psfrag{c3}{\begin{small}$d_V=0.41,\:d_A=0$\end{small}}
\psfrag{c4}{\begin{small}$d_V=0,\:d_A=0.12$\end{small}}
\psfrag{c5}{\begin{small}$d_V=0.2,\:d_A=0.3$\end{small}}
\psfrag{mtt}{\begin{Large}$\mu_{t\bar{t}}$\end{Large} (GeV)}
\psfrag{mt}{\begin{small}$2m_t$\end{small}}
{\includegraphics[width=9.cm]{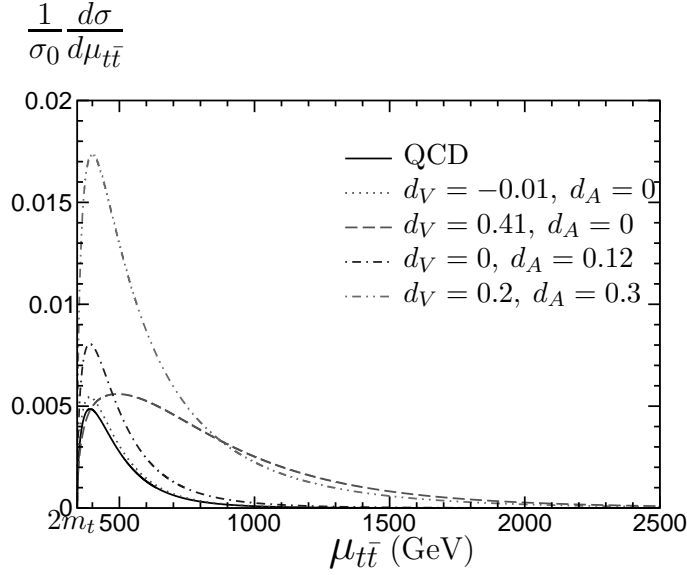}}
\caption{The $t\bar{t}$ invariant-mass distribution normalized
by $\sigma_0$: LHC energy $\sqrt{s}=10$ TeV}\label{Dism10}
\end{center}
\end{minipage}
\end{figure}

\begin{figure}[h]
\begin{minipage}{14.8cm}
\begin{center}
\psfrag{dsig}{\begin{Large}\hspace*{-0.8cm}
$\frac1{\sigma_0}\frac{d\sigma}{d\mu_{t\bar{t}}}$\end{Large}}
\psfrag{cs}{\begin{large}\hspace*{-0.0cm}$\cos\theta_t$\end{large}}
\psfrag{c1}{\begin{small}QCD\end{small}}
\psfrag{c2}{\begin{small}$d_V=-0.01,\:d_A=0$\end{small}}
\psfrag{c3}{\begin{small}$d_V=0.41,\:d_A=0$\end{small}}
\psfrag{c4}{\begin{small}$d_V=0,\:d_A=0.12$\end{small}}
\psfrag{c5}{\begin{small}$d_V=0.2,\:d_A=0.3$\end{small}}
\psfrag{mtt}{\begin{Large}$\mu_{t\bar{t}}$\end{Large} (GeV)}
\psfrag{mt}{\begin{small}$2m_t$\end{small}}
{\includegraphics[width=9.cm]{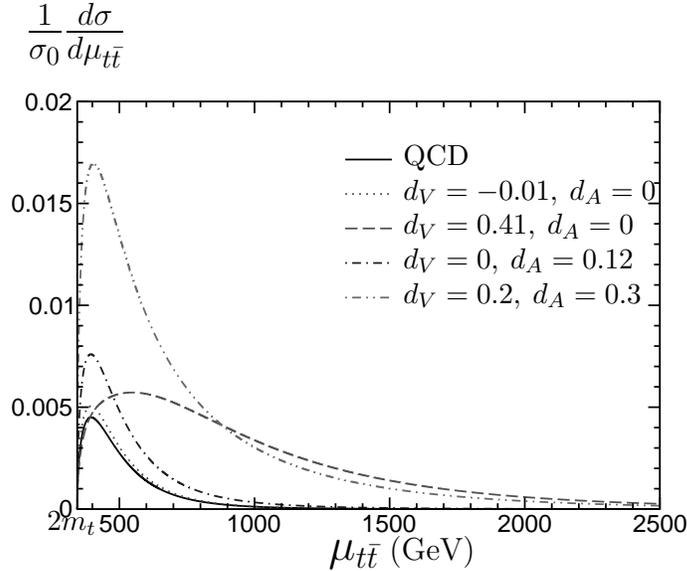}}
\caption{The $t\bar{t}$ invariant-mass distribution normalized
by $\sigma_0$: LHC energy $\sqrt{s}=14$ TeV}\label{Dism14}
\end{center}
\end{minipage}
\end{figure}

Finally, figures \ref{Dism10} and \ref{Dism14} are the $t\bar{t}$ invariant-mass
distributions $\sigma_0^{-1}d\sigma/d\mu_{t\bar{t}}$. Here again the one for
$(d_V,\,d_A)=(0.41,\,0)$ behaves a bit differently, and the others will also be
usable for our analysis except for $(d_V,\,d_A)=(-0.01,\,0)$.

It is not surprising that both the $p_{\rm T}$ and $t\bar{t}$ invariant-mass
distributions for $(d_V,\,d_A)=(0.41,\,0)$ have their peaks at a higher
$p_{\rm T}/\mu_{t\bar{t}}$ point
than the other curves. This is because $d_{V,A}$ terms can be enhanced by the
top energy as understood in ${\cal M}_{q\bar{q},gg}$, and also there occurs partial
cancellation between the $d_V$ and $d_A$ contributions when they take similar non-zero
values like $(d_V,\,d_A)=(0.2,\,0.3)$, as mentioned in the discussion of the total
cross section. This is interesting
particularly for the invariant-mass distribution: this is a mere delta-function
distribution in the parton-CM frame since $\mu_{t\bar{t}}=2E_t^*=\sqrt{\hat{s}}$ in
this frame. Therefore, we may say that we are observing in figs.\ref{Dism10} and
\ref{Dism14} the boost effects coming from the parton distribution functions, but
still the above enhancement produces some difference.

As a result, we may conclude our analyses as follows: those three differential
distributions seem to indicate that there will be some chances to observe
anomalous-coupling effects unless $|d_{V,A}|$ is very small, although their effects
are not as drastic in these quantities as in the total cross section.

\vspace*{0.5cm} % \newpage

\vspace*{0.3cm}

\vspace{0.3cm} % \newpage

% \end{center}

% 444444444444444444444444444444444444444444444444444444444444
\sec{Summary}

We have studied in this article anomalous top-gluon coupling
effects in the total cross section and several differential distributions of
$pp \to t\bar{t}X$ at LHC energies $\sqrt{s}=$ 10 and 14 TeV in the framework
of dimension-6 effective operators. We first obtained an
experimentally allowed region for the chromoelectric and chromomagnetic moments
from Tevatron (CDF/D0) data on the total cross section of $p\bar{p} \to t\bar{t}X$,
then we thereby computed the above-mentioned quantities.

We found the total cross section could get much larger than the standard-scheme
(QCD) prediction. Also the top distributions could show a different
behavior from QCD, though the non-SM effects are not as drastic 
as in the total cross section. It must be quite exciting if we actually get a huge cross
section at LHC. Conversely, if we observe cross section close to the QCD
prediction, we obtain a much stronger constraint on $d_V$ and $d_A$. In that
case, an analysis combining the Tevatron and LHC data will work very effectively.

We focused here on the top quark itself in the final state, and did not go into
detailed analyses of its various decay processes, since it would help to maximize
the number of events necessary for our studies. Indeed we have thereby shown that
there would be some chances to observe interesting phenomena. However, if we get
any nonstandard signal, we of course have to perform more systematic analyses
including decay products, i.e., leptons/$b$ quarks. We should get ready for such
an exciting situation as a next subject before actual experiments start.

\vspace{0.7cm}
% AAAAAAAAAAAAAAAAAAAAAAAAAAAAAAAAAAAAAAAAAAAAAAAAAAAAAAA
\centerline{ACKNOWLEDGMENTS}

\vspace{0.3cm}
The algebraic calculations using FORM were carried out
on the computer system at Yukawa Institute for Theoretical
Physics (YITP), Kyoto University.

\baselineskip=20pt plus 0.1pt minus 0.1pt

\vskip 0.5cm
\centerline{Note added}

After the completion of this work, we were informed that anomalous top-gluon
couplings were also studied in \cite{Lillie:2007hd} to explore the possibility
that the right-handed top quark is composite. We would like to thank Tim Tait
for calling our attention to these two papers.

\vspace*{0.8cm}
% RRRRRRRRRRRRRRRRRRRRRRRRRRRRRRRRRRR

\end{document}